\documentclass[aps,pre,twocolumn,showpacs]{revtex4}
\usepackage[]{graphicx}

\begin{document}

\title{Linear local modes induced by intrinsic localized modes in a
monatomic chain}
\author{V. Hizhnyakov}
\author{ A. Shelkan}
\email{shell@fi.tartu.ee}
\affiliation{Institute of Physics, University
of Tartu, Riia 142, 51014 Tartu, Estonia}
\author{M. Klopov}
\affiliation{Institute of Physics, Tallinn University of
Technology, Ehitajate 5, 19086 Tallinn, Estonia}

\author{S. A. Kiselev}
\author{A. J. Sievers}
\affiliation{Laboratory of Atomic and Solid
State Physics, Cornell University, Ithaca, New York
14853-2501,USA}

\begin{abstract}

A theory is developed to describe the effect of an intrinsic localized
mode (ILM) on small vibrations in a monatomic chain with hard
quartic anharmonicity. One prediction is the appearance in the
chain of linear local modes nearby the ILM. To check this result,
MD calculations of vibrations under strong local excitation are
carried through with high precision. The results fully confirm the
prediction.
\end{abstract}

\maketitle

\draft

\section{Introduction}

The realization \cite{dolg,sievtak,taken} that there can exist
stable strongly localized excitations in perfect classical
anharmonic lattices has led to theoretical studies exploring a
variety of possibilities
\cite{Page,KisBick,SievPage,flach,LaiSi,KisBS,Schrod,KisSiev,Bishop}.
The frequency of such a localized vibration depends on its
amplitude and lies outside the phonon spectrum. Such excitations
are called intrinsic localized modes (ILMs) \cite{sievtak}, to
emphasize their similarity in appearance to defect impurity modes,
or called discrete breathers \cite{flach} or discrete solitons
\cite{Toda,christ} to make a connection to solitons in continuous
systems. Some experimental evidence for the existence of intrinsic
localized modes in 1-D lattices in microscopic and macroscopic
lattices has been demonstrated
\cite{Swanson,Schwartz,TrMaOr,Bind,Craig,Fleisch,MSato,Wrubel}.

%

It is to be expected that the appearance of an ILM would change
the local properties of the lattice including the local phonon
dynamics. This back reaction on the phonon spectrum should have
physical consequences since the ILM could induce local
modes outside the plane wave spectrum.
The presence of these additional resonances and their
dependence on the amplitude of the ILM should add some complexity
to the energy relaxation rate of the ILM to the phonon bath.  In a different
direction, since an ILM can move through the lattice
it is expected that such trapped local modes would also move or, more likely,
tend to inhibit the translational motion of such localized energy.

In this communication we examine analytically the small amplitude
vibrations of a 1-D  nonlinear chain with intersite coupling in
the presence of an ILM.  The results show that an ILM with
sufficient nonlinear amplitude stabilizes the appearance of
linear local modes (LLMs) above the top of the plane wave spectrum.
Next molecular dynamics (MD) simulations are used to verify these
findings. First an ILM is formed and the power spectrum
calculated, next the initial conditions are changed and very small
amplitude shifts associated with a LLM are added to the neighboring
atoms and the power spectrum calculated again.
The cases when the new spectra contain
one additional weak peak above the phonon spectrum are studied in
detail. The frequencies of the additional peaks do not depend
on their amplitudes but they do depend on the ILM amplitude
consistent with the linear requirement.  Good agreement is
obtained between the positions of the additional peaks and the
theoretically calculated frequencies of the LLMs.

\section{Theory}

Let us consider an anharmonic monatomic chain with interactions of
nearest-neighboring atoms. Taking into account only quartic
anharmonicity, the equation of motion of atoms in this model reads
\begin{equation}
\ddot{U}_n= \sum_{n' = n \pm 1}\big[k_2( U_{n'}-U_n) + k_4(
U_{n'}-U_n)^3 \big], \label{difeq}
\end{equation}
where ${U}_{n}$ are the reduced displacements of atoms located at
the site $n$ of the chain, the subscripts $n$ indicate the number
of the site, $k_2$ and $k_4$ are the parameters of harmonic and
anharmonic springs, whereat $k_2 = \omega_m^2/4 $ determines the
top phonon frequency $\omega_m$. If $k_4>0$ then  ILMs may be
excited in the chain with the frequencies above the allowed phonon
spectrum \cite{sievtak,Page}.


To describe small vibrations of the chain in presence of an ILM,
we add to $U_n$ an infinitesimal displacement $q_n$: $ U'_n (t) =
U_n (t) + q_n (t)$. For the displacements of the ILM we take
$U_n(t)=A_n \cos{\omega_L t}$, where $A_n$ is the amplitude
parameter of the ILM which may be both, positive or negative (the
small contribution of higher-order harmonics is omitted). The
shift $U'_n (t)$ also satisfies Eq. (\ref{difeq}). Subtracting
from this equation the equation for $U_n(t)$ we get
\begin{equation}
\ddot{q}_n = \sum_{n' = n \pm 1}\big[k_2+ 3 k_4 (A_n -A_{n'})^2
\cos^2{\omega_L t}\big](q_{n'}-q_n). \label{Eq_q_n}
\end{equation}
One \ can \ divide \ $q_n$ \ into two parts: \ 1) the shifts \ $q_{0,n}$ \ 
describing small variations of the ILM  (they have been considered
in \cite{HNS,SheHiKlo}) \ 2) all other shifts \ $q_{1,n}$. Here we are
interested in the stable solutions of equation (\ref{Eq_q_n}) which
correspond to the latter modes. These modes  are orthogonal to (i.e.
independent from) the ILM.  We can take the orthogonality condition
into account if we add in Eq. (\ref{Eq_q_n}) the  term \ $ \lambda
\sum_{n'} A_n A_{n'} q_{n'},$ where the Lagrange multiplier \ 
$\lambda$ \  should \ be found  from the orthogonality condition of the
LLM under consideration. However, if one considers the linear modes
of symmetry different from the ILM then there is no need to add the
term $\propto \lambda$, since the orthogonality conditions are
already fulfilled.

To identify LLMs, we first apply the
rotating wave approximation which corresponds to the replacing in
equation (\ref{Eq_q_n}) the factor \ $\cos^2{\omega_L t} = (1+ \cos{2
\omega_L t})/2$ \ by \ $1/2$ \ (i.e. we neglect the oscillating in time
term $(1/2)\cos{2 \omega_L t}$). In this approximation  the equation
of motion of the linear modes takes the form:
\begin{equation}
-\ddot{q}_n =\sum_{n'}(D_{n n'} + v_{n n'})q_{n'}, \label{qddot}
\end{equation}
\noindent where $D_{n n'}=k_2 (2 \delta_{n n'} -\delta_{n \pm
1,n'})$ is the dynamical matrix of the perfect monatomic chain,
\begin{eqnarray}
v_{nn'} & = & \frac{3k_4}{2} \big[ \delta_{n n'}
\big((A_n-A_{n+1})^2+
(A_n-A_{n-1})^2 \big)  \nonumber\\
 & - &\delta_{n \pm 1,n'}(A_n-A_{n'})^2 \big] + \lambda A_n A_{n'}
 \label{matrix_v}
\end{eqnarray}
is the perturbation of the dynamical matrix,

The effect of the perturbation can be found by the Lifshitz
method \cite{maradudin}. Thus a linear local mode $l$ exists if
the imaginary part of the Green's function of the perturbed
lattice has a pole at $\omega_l$ outside the allowed phonon
spectrum. The latter function can be found from the equation
\begin{equation}
G (\omega) = \big(I-G^{(0)} (\omega)\, v \big)^{-1} G^{(0)}
(\omega), \label{Green_omega}
\end{equation}
where $G$, $v$, and $G^{(0)}$ are matrices; $v$ is given by Eq.
(\ref{matrix_v}),
\begin{equation}
G^{(0)}_{nn'}(\omega) = [-\rho(\omega)]^{|n-n'|}/ \omega
\sqrt{\omega^2 -1}, \label{cheneGreen}
\end{equation}
is the $(n,n')$-component of the Green's function matrix of the
perfect chain \cite{economou}, $\rho (\omega) = (\omega -
\sqrt{\omega^2-1})^2 \leq 1$ (the units $\omega_{m} = 1$ are used).
The amplitude parameters of a LLM are given by the relation
\begin{equation}
a_{n,\,l}= G_{nn_1}(\omega_l)/G_{n_1n_1}(\omega_l). \label{norm_an}
\end{equation}


The time oscillatory terms in Eq. (\ref{Eq_q_n}), neglected in the rotating wave
approximation, can lead to new effects not found in the vibrations
of a lattice with a static defect. This difference follows from the
Floquet theorem according to which in a periodically time-dependent
system with the period $T$, besides the excitations with the
frequency $\omega_l$, there exist also excitations with the
frequencies $|\pm \omega_l + 2 \pi N /T|$, where $N$ is an integer.
In our case $T= \pi/\omega_L$; the repetitions of the frequency
result from the time-dependent part of the anharmonic interaction of
the LLMs with ILMs.

The \ time \ oscillatory terms \ in \ Eq. (\ref{Eq_q_n}) \ can \ also \ cause 
a \ renormalization \ of \ the frequencies \ of \ linear modes. This \ 
follows \ from \ the \ fact \ that \ these \ terms \ oscillate \ in \ time \ with \ the
frequencies \ $2\omega_L- \omega$ \ and \ $2 \omega_L + \omega$ \ (here
$\omega$ is the frequency of a small vibration), one of which,  $2
\omega_L - \omega_l,$ may be comparable with the  perturbation $v$.
As a result, the frequencies of some linear modes may acquire complex
values, which will cause these modes to become
unstable \cite{Kiv,SadPage,Daumont,LaiSievers}.

To find whether a LLM is stable or not we apply in  Eq.
(\ref{Eq_q_n}) the expansion $q_n=\sum_j e_{nj}x_j$, where $e_{n j}$
is the normalized contribution of the atom $n$ to the normal
coordinate $x_j$ of the perturbed lattice. We get
\begin{equation}
-\ddot{x}_{l}= \omega_l^2 x_{j} +  \cos{2 \omega_{L} t} \,\sum_{n
n'} \sum_{j} e_{n l} \, w_{n n'} e_{n' j} \, x_{j}. \label{ddotxl}
\end{equation}
The effect of the time oscillatory term in Eq. (\ref{ddotxl}) is
most important for the  modes with frequencies close to $\omega_L$.
Supposing that only the LLM under consideration has such a
frequency, we can neglect in the right-hand-side of this equation
the terms $j \neq l$. In this approximation
\begin{equation}
-\ddot{x}_{l}= \omega_l^2(1+h \cos{2\omega_L t}) \, x_l.
\label{parametric}
\end{equation}
where $ h= \omega_l^{-2} \,\sum_{n n'} e_{n l} \, w_{n n'}\, e_{n'
l}.$ (Here the inhomogeneous term $\propto x_L (t) \cos{2
\omega_{L} t}$ also has been neglected; this term describes the
infinitesimal forced oscillations with the frequencies $\omega_L$
and $3 \omega_L$resulting in the infinitesimal shift of the ILM
phase.)
Equation (\ref{parametric}) is the Mathieu equation describing the
parametric resonance \cite{Landau}. The term $\propto h$
 may result in a) renormalization of the frequency of the LLM,
or b) in the modes instability; in the latter case the renormalized
frequency is complex. Below we consider the renormalization and the
instability conditions numerically.

\subsection{LLMs induced by even ILM}

Both even and odd ILMs can exist in the nonlinear
monatomic chain \cite{Page}. Here we consider an even ILM to be the effective
defect in this lattice calculation, since this mode is stable with
respect to small translational fluctuations \cite{SadPage}. This
choice will permit calculations of ILMs + LLMs with high precision.
The amplitude patterns for such an ILM for different values of the
dimensionless nonlinear parameter, $k_4 A_0^2/k_2$ , are given in
rows 3 and 4 of Table 1.

An even ILM can induce the appearance of both, odd and even LLMs. In
the case of odd LLMs  the $\lambda$- multiplier in Eq.
(\ref{matrix_v}) equals zero. Then, if one considers an ILM
localized of six neighboring central atoms, the perturbation matrix
$v$ (see Eq. (\ref{matrix_v})) is the six-range symmetric matrix
with the following nonzero elements:
\begin{eqnarray}
v_{00}&=& v_{11}=  \gamma_0 +\gamma_1, \,\,v_{01}=-\gamma_0 ,
\,\,v_{12}=-\gamma_1, \nonumber\\
v_{22}&=&\gamma_1+\gamma_2,\ v_{23} = - \gamma_2, \ v_{33}=\gamma_3,
\label{matrix}
\end{eqnarray}
and with few matrix elements with negative index satisfying the
relations $v_{-n-n'} =v_{n+1 n'+1}$ ($n > 0$) . Here $\gamma_n
\approx 3 k_4 (A_n-A_{n+1})^2/2 k_2$ are the changes of the springs
(in dimensionless units).
The effect of an even ILM upon even phonons is described by the
perturbation matrix given by Eq. (\ref{matrix_v}), which is equal to
the previous perturbation matrix $v$ plus the term $\lambda A_n
A_{n'}$; the value of $\lambda$ must be found self-consistently from
the orthogonality condition $\sum _n e_{n,evn}A_n =0$.

Inserting the matrix  $v$ into Eqs. (\ref{Green_omega}) and
(\ref{norm_an}) gives the frequencies and relative amplitudes of the
LLMs.
The parametric resonance parameter $h$, which determines the
frequency correction equals
\begin{equation}
h=\frac{3 K_4 A_0^2}{K_2 \omega_l^2} \sum_{n \geq 1} \left(
|e_{n}|+|e_{n+1}| \right)^2 \left(|A_n|+|A_{n+1}| \right)^2.
\label{hfin}
\end{equation}

\vspace{0mm} {\bf Table 1.} Analytical calculations of odd and even
LLMs for a lattice containing an even ILM with a range of nonlinear
parameter $k_4 A_0^2/k_2$ values. Given are the relative ILM
frequency $\omega_L/\omega_m$, the ILM amplitudes of the four
next-to central particles $A_n$ with $A_{-n} = -A_{n+1}$, the
relative frequency of the odd and even LLM $\omega_{odd}/\omega_m$
and $\omega_{evn}/\omega_m$, the amplitudes of these modes
$a_{-n,odd} = a_{n+1,odd}$ and $a_{-n,evn} = -a_{n+1,evn}$ ($n \ge
0$); $A_1-A_0 =1$; $a_{1,odd}=a_{1,evn} =1.$ (The frequencies of the
odd LLMs are corrected for the time oscillatory perturbation; $h$
and $h_{cr}$ are the corresponding parameters of this perturbation.)

\begin{center}
\begin{tabular} {|c|c|c|c|c|c|c|c|}
\hline
\vspace{-4mm} & & & & & & & \\
$k_4 A_0^2/k_2$& 0.50 & 0.75 & 1.00 & 1.25 &1.50 & 1.75 & 2.00 \\
\vspace{-4mm} & & & & & & & \\
\hline
\vspace{-4mm} & & & & & & & \\
\vspace{-4mm} $\omega_L/\omega_m$& 1.36 & 1.51 & 1.64 & 1.77 & 1.88 & 2.00 & 2.10 \\
 & & & & & & & \\
\hline
\vspace{-4mm} & & & & & & & \\
$-A_2$& 0.199 &0.170 &0.153 &0.141 &0.133 &0.127 &0.121 \\
\vspace{-4mm} & & & & & & & \\
\hline
\vspace{-4mm} & & & & & & & \\
\vspace{-4mm} $A_3$ & 0.042 & 0.027 & 0.019 & 0.015 & 0.012& 0.010& 0.008 \\
 & & & & & & & \\
\hline \hline
\vspace{-4mm} & & & & & & & \\
$\omega_{odd}/\omega_m$& 1.288 & 1.418 & 1.486 & 1.58& 1.677 & 1.78 & 1.86 \\
\vspace{-4mm} & & & & & & & \\
\hline
\vspace{-4mm} & & & & & & & \\
$a_{1,odd}$ & 0.729 & 0.784 & 0.819 & 0.844 & 0.862 & 0.876& 0.887 \\
\vspace{-4mm} & & & & & & & \\
\hline
\vspace{-4mm} & & & & & & & \\
\vspace{-4mm} $a_{3,odd}$ & 0.347 &0.262 &0.212 &0.179 &0.156 & 0.138 &0.124 \\
 & & & & & & & \\
\hline
\vspace{-4mm} & & & & & & & \\
\vspace{-4mm} $h_{odd}$ & 0.456 &0.560 &0.629 &0.679 &0.711 & 0.746 &0.770 \\
 & & & & & & & \\
\hline
\vspace{-4mm} & & & & & & & \\
\vspace{-4mm} $h_{odd,cr}$ & 0.52 &0.63 &0.77 &0.84 &0.89 & 0.94 &0.98  \\
 & & & & & & & \\
\hline \hline
\vspace{-4mm} & & & & & & & \\
$\omega_{evn}/\omega_m$ & 1.009 &1.031 &1.063 &1.102&1.142& 1.179& 1.219 \\
\vspace{-4mm} & & & & & & & \\
\hline
\vspace{-4mm} & & & & & & & \\
$a_{1,evn}$ & 0.383& 0.308& 0.273 &0.252 &0.238& 0.228& 0.220 \\
\vspace{-4mm} & & & & & & & \\
\hline
\vspace{-4mm} & & & & & & & \\
\vspace{-4mm} $-a_{3,evn}$ &1.143 & 0.822 &0.639 &0.522 &0.442 &0.384&0.340 \\
 & & & & & & & \\
\hline
\vspace{-4mm} & & & & & & & \\
\vspace{-4mm} $\tilde{\lambda} \cdot 10^2$ & 0.70& 0.83& 0.87
&0.92
 & 0.96 &0.98  &1.01 \\
 & & & & & & & \\
\hline

\end{tabular}

\end{center}

In Table 1 the calculated values for odd and even LLMs are
presented for different values of the nonlinear parameter of the
ILM, $k_4 A_0^2/k_2$. Comparing the LLM amplitude patterns with the
amplitudes of the ILM shows that they are mutually orthogonal and
the ILM and the LLM belong to the different degrees of freedom.
Note that the frequency of the odd LLM is rather close to $\omega_L$. For
this mode consideration of the time oscillatory term in Eq.
(\ref{parametric}) is important. The numerical calculations of the
Mathieu equation (\ref{parametric}) shows that in all cases the
parametric resonance parameter $h$ is less than its critical value
$h_{cr}$ for the instability. The frequency of the even LLM is
significantly different from $\omega_L$ so that the correction caused by the
time oscillatory term is now small.

\section{MD simulations}

To perform MD simulations of ILMs and LLMs, we integrate numerically
the equation of motion given by Eq. (\ref{difeq}). We calculate the
vibrational amplitudes  and velocities of atoms of the finite chain
with $10^4$ atoms in $10^5$ time points. The time interval includes
$\sim 10^3$ periods of the ILMs. We also calculate the vibrational
power spectrum. To decrease the background in the spectrum, Fourier
transformations are carried out between the time points
corresponding to the same phase of vibrations.
For the initial condition, we used the ILM displacements of the six central
atoms from their equilibrium positions.
The amplitude difference of the two central
atoms is fixed at unity with $k_2=100$ ($\omega_{m} = 20$).
In this case the actual values of $k_4$ for ILMs with frequency
$\leq 2\omega_m$ are of the order of $k_2$.
The initial shifts of other four central atoms are selected so
that only the ILM is excited.  The resulting amplitude parameters $A_n$ of
six central atoms are those shown in Table 1.

\begin{figure}[th]
\includegraphics[angle=-90,width=.49\textwidth]{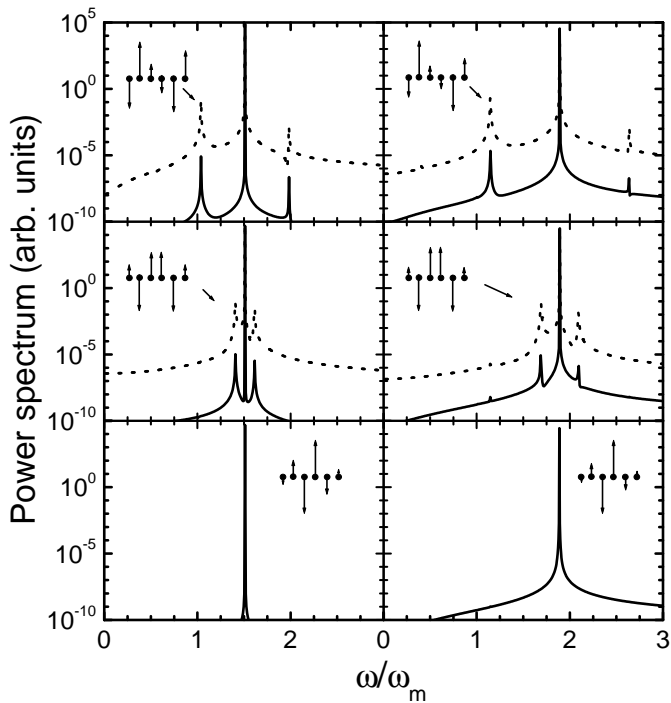}\hspace*{0em}
\caption{Power spectra of ILMs for two different nonlinear parameter
values with and without LLMs. Left column of figures: the main
spectral feature is an even ILM with $k_4 A_0^2/k_2$ = 0.75, right
column of figures: same even ILM but with $k_4 A_0^2/k_2$ = 1.5.
Bottom panels: the unperturbed even ILM; middle panels: an ILM
perturbed by an odd LLM; top panels: an ILM perturbed by an even
LLM. Although the dotted and solid lines differ by $10^2$ in the LLM
amplitude its frequency remains unchanged. The amplitude patterns of
the ILMs and LLMs are shown by arrows.}
\end{figure}

Examples of the calculated power spectrum of an even ILM are given
in the bottom panels of Fig. 1. Two different values of the
nonlinear parameter are shown. Each \ spectrum \ has \ a single \ peak \ 
with \ a \ frequency $\omega_L > \omega_m$. Not shown are the much weaker
peaks associated with the higher odd harmonics at $3 \omega_L$, $5
\omega_L$, etc. Hence the bottom panels demonstrate that our initial
conditions correspond solely to the excitation of an ILM (at least
with the accuracy $10^{-14}$).

When small additional amplitude shifts appropriate to the odd or
even symmetry LLM presented in Table 1 are added to the ILM
amplitude pattern, then additional small \ peaks \ appear \ 
in \ the \ power \ spectrum, \ one \ at $\omega < \omega_L$, and an even smaller
symmetrically situated peak at $\omega > \omega_L$. The middle panels
in Fig. 1 show the case for an odd LLM, with its amplitude pattern,
and the top panel for an even LLM, again with its amplitude pattern.
One finds analogous sideband spectra for other  $k_4A_0^2/k_2$
values. These data show that the frequencies of the new spectral
peaks depend on the amplitude and hence the frequency of the ILM. At
the same time the solid and dotted traces presented in Fig. 1, where
the LLM amplitude is varied by $10^2$, demonstrate that these LLM
frequencies do not depend on their own amplitudes as expected for a
linear response (the amplitudes of the LLMs are $\sim 10^5$ and
$\sim 10^3$ times less than the amplitude of the ILM). The
$\tilde{\omega}_l = 2 \Omega_L - \omega_l$ of the symmetrically
situated weak peak indicates that this spectral feature is the four
wave mixing response of the LLM with the  ILM.

\section{Discussion and Conclusions}

There \ is \ value \ in \ a \ quantitative \ comparison \ of \ the \ analytical \ 
results \ and \ MD \ simulations \ for \ the \ LLMs. Figure 2 presents the frequencies of
the LLMs obtained by both these methods.
\begin{figure}[th]
\includegraphics[angle=-90,width=.50\textwidth]{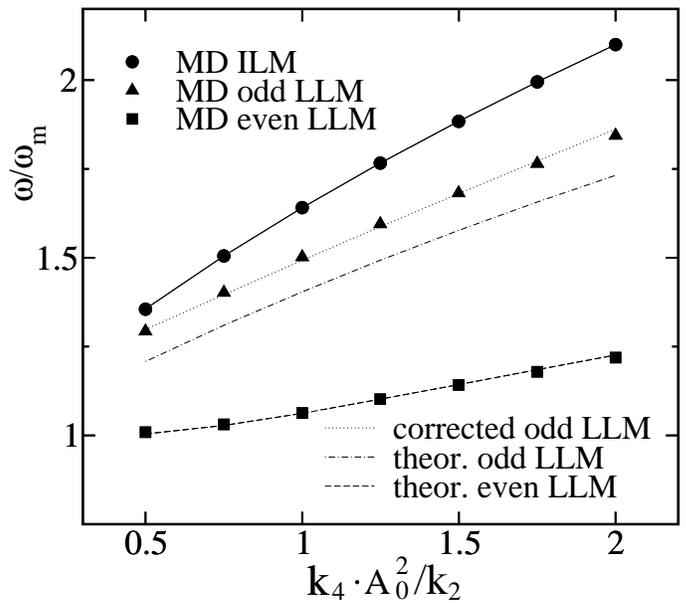}\hspace*{0em}
\caption{The dependence of the frequencies of the odd and even LLMs
on the dimensionless nonlinear parameter $k_4 A_0^2/k_2$ of the ILM.
The frequencies of the odd LLM are given with (dotted line) and
without (dash-dotted line) taking into account the time-oscillatory
term $h \cos{2\omega_l t}$ in Eq. (\ref{parametric}). The ILM
frequency dependence is also shown.}
\end{figure}
Inspection of these results shows that there is excellent agreement
between the two methods. Note that the time oscillatory terms  provide
a significant correction for odd LLM case but are insignificant for the
even LLM. The reason is \ in the relatively small difference of \ 
$\omega_L-\omega_{odd}$ \ as compared to $\omega_L-\omega_{evn}$.

We have presented a theory which allows one to describe the effect
of an ILM on phonons in a nonlinear monatomic chain. The
prediction of linear localized modes above the top of the plane wave
spectrum is in good agreement with MD simulations. Basically
the appearance of the ILM changes the nearby nonlinear spring
constants sufficiently so that linear local modes can also
appear. The resulting lattice
perturbation produced by the ILM is analogous to that associated
with a force constant defect in a linear lattice in that a small number of
degrees of freedom can be strongly perturbed but all phonons are
perturbed to some extent. Looking ahead to more realistic
diatomic lattices involving two body potentials \cite{KisBS},
we expect that a variety of LLM possibilities may appear.
These could include modes of the local, gap, and resonant types
as well as tunneling states, all made possible by the
fundamental combination of nonlinearity and lattice discretness.
One may anticipate that, in analogy with the
previously studied force constant defect cases,
ILM-induced IR and Raman-activity may occur
throughout the entire phonon spectrum \cite{Barker,Cardona}.

There are also some very interesting differences between
the properties of extrinsic and intrinsic localized excitations evident
from our study of this simple model system. One is
that an ILM introduces a periodically time-dependent perturbation
that supports nonlinear mixing processes between the ILM
and the LLMs and a second is that for a
moving ILM the "effective defect space" with its associated
LLMs will travel together with the nonlinear excitation.

\section{Acknowledgments}
This research is supported by the Estonian Science Foundation,
Grant No 6534, the U.S. National Research Council Twinning Program
with Estonia, by NSF-DMR under Grant No. 0301035 and by the
Department of Energy under Grant No. DE-FG02-04ER46154.

\end{document}